\documentclass[aps,pra,twocolumn,amsmath,amssymb,footinbib,showpacs,longbibliography]{revtex4-1}

\newcommand{\Jnature}{Nature (London)}
\newcommand{\Jnatphys}{Nat. Phys.}

\newcommand{\Jprl}{Phys. Rev. Lett.}
\newcommand{\Jpr}{Phys. Rev.}
\newcommand{\Jpra}{Phys. Rev. A}
\newcommand{\Jprb}{Phys. Rev. B}

\newcommand{\Jrmp}{Rev. Mod. Phys.}

\newcommand{\Jepl}{Europhys. Lett.}
\newcommand{\Jnjp}{New J. Phys.}

\newcommand{\Jadvphys}{Adv. Phys.}
\newcommand{\Jannphys}{Ann. Phys. (NY)}
\newcommand{\JAnnualRevCondMat}{Annual Rev. Cond. Mat. Phys.}

\newcommand{\JCommMathPhys}{Comm. Math. Phys.}

\usepackage{times}
\usepackage[english]{babel}
\usepackage{latexsym}
\usepackage{graphics}
\usepackage{subfigure}
\usepackage{epsfig}
\usepackage{color}
\usepackage{hyperref}
\usepackage{braket} 
\usepackage[T1]{fontenc}
\usepackage[latin9]{inputenc}
\usepackage{amstext}
\usepackage{amssymb}
\usepackage{graphicx}
\usepackage{esint}
\usepackage{braket}
\usepackage{babel}
\usepackage{amsmath}
\hypersetup{
colorlinks=true,
citecolor=blue,
linkcolor=red,
urlcolor=black
}

\definecolor{darkred}{rgb}{0.5,0,0}
\definecolor{darkgreen}{rgb}{0,0.4,0}

\definecolor{orange}{rgb}{1,0.5,0}

\newcommand{\e}{\textrm{e}}
\newcommand{\ie}{\textit{i.e.}}

\newcommand{\qav}[1]{{\overline{#1}}}
\newcommand{\Ucrit}{u_{\textrm{c}}}
\newcommand{\gammacrit}{\gamma_{\textrm{c}}}

\newcommand{\Vs}{V_{\textrm{s}}}
\newcommand{\Vg}{V_{\textrm{g}}}
\newcommand{\Vphi}{V_{\varphi}}
\newcommand{\tildeVphi}{\tilde{V}_{\varphi}}
\newcommand{\tildeVg}{\tilde{V}_{\mathrm{g}}}
\newcommand{\VCE}{V_{\textrm{\tiny CE}}}
\newcommand{\Vm}{V_{\textrm{m}}}
\newcommand{\ksp}{k_\mathrm{sp}}

\newcommand{\lettersection}[1]{\paragraph*{#1.---}}
\renewcommand{\acknowledgments}{\bigskip\noindent}


\begin{document}

\newcommand{\ttitle}{Twofold correlation spreading in a strongly correlated lattice Bose gas}
\title{\ttitle}

\author{Julien Despres}
\affiliation{
 CPHT, Ecole Polytechnique, CNRS, Universit\'e Paris-Saclay, F-91128 Palaiseau, France
}

\author{Louis Villa}
\affiliation{
 CPHT, Ecole Polytechnique, CNRS, Universit\'e Paris-Saclay, F-91128 Palaiseau, France
}

\author{Laurent Sanchez-Palencia}
\affiliation{
 CPHT, Ecole Polytechnique, CNRS, Universit\'e Paris-Saclay, F-91128 Palaiseau, France
}

\date{\today}

\begin{abstract}
We study the spreading of correlations in the Bose-Hubbard chain, using the time-dependent matrix-product
state approach. In both the superfluid and the Mott-insulator phases, we find that the time-dependent correlation functions generally display a universal twofold cone structure characterized by two distinct velocities. The latter are related to different microscopic properties of the system and provide useful information on the excitation spectrum. The twofold spreading of correlations has 
profound implications on experimental observations that are discussed.
\end{abstract}

\maketitle

In the last decades, simultaneous progress of the many-body quantum theory and the experimental control of quantum matter in condensed matter and atomic, molecular, and optical physics has given dramatic momentum to the understanding of the out-of-equilibrium dynamics of correlated quantum systems~\cite{polkovnikov2011,eisert2015,*pekola2015,langen2015,lewenstein2007,*bloch2008, NaturePhysicsInsight2012bloch,*NaturePhysicsInsight2012blatt,*NaturePhysicsInsight2012aspuru-guzik,*NaturePhysicsInsight2012houck}.
The spreading of quantum correlations governs many fundamental phenomena, including the propagation of information and entanglement, thermalization, and the area laws for entanglement entropy. For lattice systems with local interactions, 
the existence of Lieb-Robinson (LR) bounds implies the emergence of a causal light cone beyond which the correlations are exponentially 
suppressed ~\cite{lieb1972,bravyi2006,hastings2006}.
So far, light-cone-like spreading of correlations has been reported in short-range
interacting models~\cite{barmettler2012,cheneau2012,carleo2014,manmana2009} as well as long-range models~\cite{jurcevic2014, richerme2014, hauke2013, eisert2013, cevolani2015, schachenmayer2015b, buyskikh2016, cevolani2016, frerot2018, cevolani2018}
where weaker LR bounds exist~\cite{hastings2006,foss-feig2015}.
However, many questions remain open. For instance, it is still debated whether a non-linear cone emerges in generic long-range systems, for which different results point towards either super-ballistic, ballistic or sub-ballistic spreading.
It was recently proposed that these apparently conflicting results can be reconciled by the coexistence of several signals governed by different scaling laws~\cite{cevolani2018}.
This behavior may be related to the non-linearity of the quasiparticle excitation spectrum, and may also appear in systems with short-range interactions.
In the later case, it is expected that both the signals spread ballistically but with different velocities.
However, this picture relies on meanfield theory, which
ignores potentially important dynamical effects, such as quasiparticle collisions and finite lifetime.

In this work, using an exact many-body approach beyond meanfield theory, we demonstrate the emergence of a universal twofold dynamics in the spreading of correlations for a generic short-range, strongly correlated quantum model.
Specifically, we consider the one-dimensional Bose-Hubbard model and use time-dependent tensor network techniques based on matrix product states.
Spanning the phase diagram, we almost always find a twofold structure of the space-time correlation pattern, characterized by two distinct velocities, essentially irrespective of the correlation function.
Exceptions, discussed below, only appear for particular cases.
In the superfluid meanfield regime and in the Mott insulator phase, the two spreading velocities are readily interpreted from the properties of the corresponding excitation spectra, which are known.
In the strongly correlated superfluid regime, only the sound velocity is known.
There, our results show beyond Luttinger liquid behavior and provide useful information about the excitation spectrum beyond the phonon branch.
The emergence of a universal twofold spreading of correlations has profound implications on 
experimental observations, which we discuss, including with a view towards extensions to long-range systems.

\begin{figure}[b!]
\includegraphics[scale = 0.27]{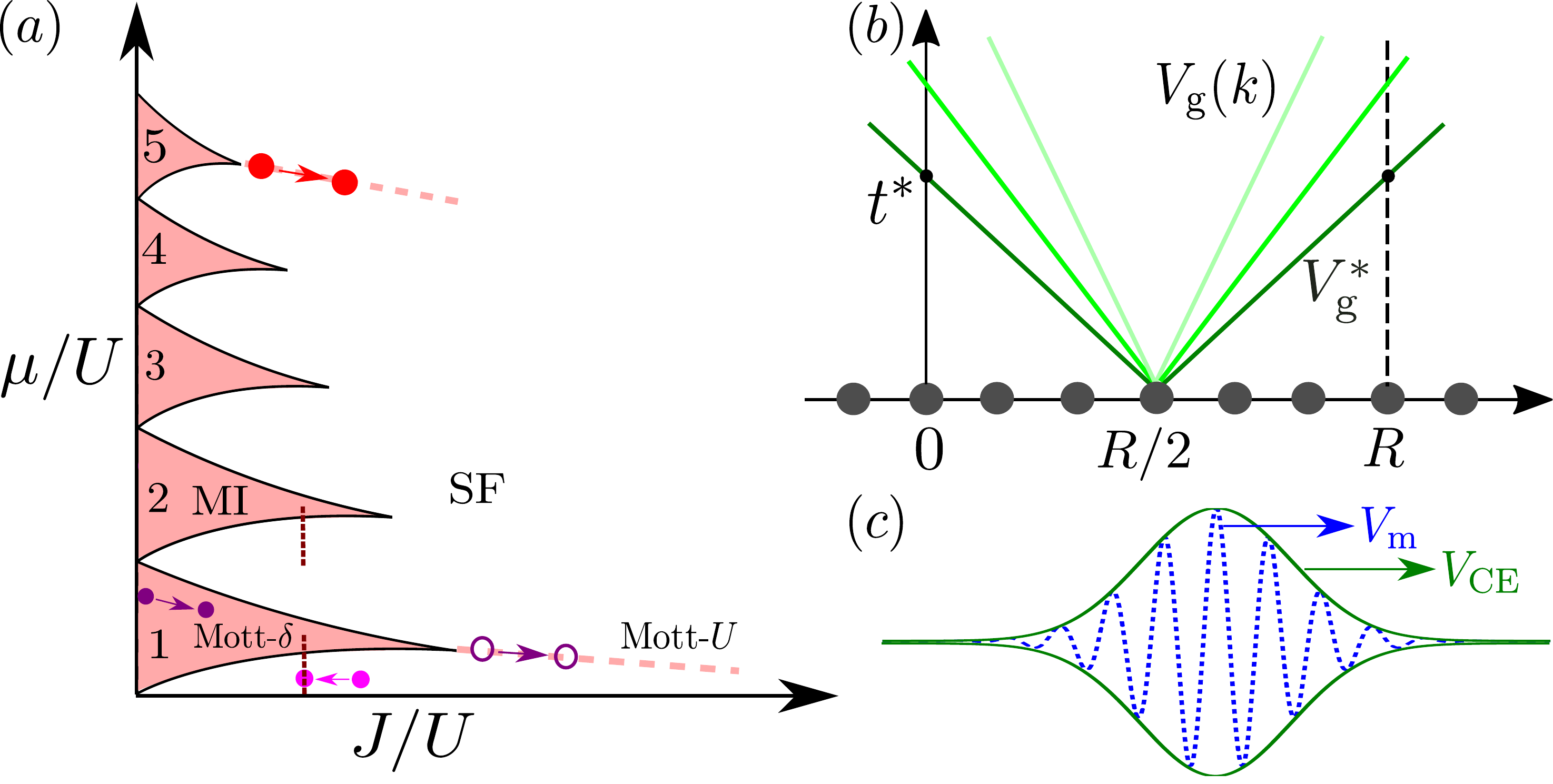}
\caption{\label{fig:phase_diagram}
Quantum quench in the Bose-Hubbard model.
(a)~Schematic phase diagram as a function of the inverse interaction strength and chemical potential, comprising a MI phase (pink lobes at integer fillings $\bar{n}$) and a SF phase.
The Mott-$U$ transition at unit filling is indicated by the dashed pink line and the Mott-$\delta$ transition by the vertical line.
The arrows indicate the various quenches considered in this work.
(b)~Generation of correlations between two points at a distance $R$ by pairs of counter-propagating quasiparticles emitted at the mid-point $R/2$. The first correlation is generated by the fastest quasiparticles at the activation time $t^\star=R/2\Vg^\star$.
(c)~Correlation spreading in the vicinity of the correlation edge (CE). The correlation function forms a periodic series of maxima moving at the velocity $\Vm = 2 V_\varphi^\star$, with an envelope moving at the velocity $\VCE = 2\Vg^*$.
}
\end{figure}

\lettersection{Model and approach}\label{sec:BHm}
The Hamiltonian of the one-dimensional (1D) Bose-Hubbard (BH) model, considered throughout this work, reads as

\begin{equation}\label{BHm}
\hat{H}=-J \sum_{R} \left( \hat{a}^{\dagger}_{R} \hat{a}_{R+1}+\mathrm{h.c.}\right)+\frac{U}{2}\sum_R\hat{n}_{R}(\hat{n}_{R}-1),
\end{equation}

where $\hat{a}_R$ and $\hat{a}_{R}^{\dagger}$ are the bosonic annihilation and creation operators on site $R$,
$\hat{n}_R = \hat{a}^{\dagger}_R \hat{a}_R$ is the occupation number (filling),
$J$ is the hopping amplitude, $U>0$ is the repulsive on-site interaction energy, and
the lattice spacing is fixed to unity ($R\in\mathbb{Z}$).
At equilibrium and zero-temperature, the phase diagram of the 1D BH model is well known ~\cite{sachdev2001,cazalilla2011}, and
sketched on Fig.~\ref{fig:phase_diagram}(a).
It comprises a superfluid (SF) and a Mott insulator (MI) phase, determined by the competition of the hopping, the interactions, and the average filling $\qav{n}$ (or, equivalently, the chemical potential $\mu$).
For commensurate filling, $\qav{n}\in\mathbb{N}^*$ the SF-MI (Mott-$U$) transition is of the Berezinskii-Kosterlitz-Thouless type, at the critical value $\Ucrit \simeq 3.3$ for unit filling
($\qav{n}=1$) in 1D~\cite{kuhner2000,kashurnikov1996exact,ejima2011,rombouts2006}.
For incommensurate filling, the Bose gas is a SF for any value of $U/J$.
The commensurate-incommensurate (Mott-$\delta$) transition, of the meanfield type, is then driven by doping when $\qav{n}$ approaches a positive integer value for sufficiently strong interactions.

We study the out-of-equilibrium dynamics of the BH model by applying a sudden global quench~\cite{calabrese2006,barmettler2012,kollath2007,moeckel2008,manmana2009,roux2010,navez2010,carleo2014,krutitsky2014},
as can be realized in ultracold-atom experiments~\cite{greiner2002b,cheneau2012,langen2013,geiger2014}.
We start from the ground state for some initial value of the interaction parameter $(U/J)_0$
and let the system evolve with a different value of $U/J$.
In the following, we consider a variety of quenches, spanning the phase diagram, see arrows on Fig.~\ref{fig:phase_diagram}(a).
We study the spreading of both phase and density fluctuations, via the connected correlation functions
$G_1(R,t) = \langle \hat{a}^\dagger_R (t) \hat{a}_0 (t) \rangle - \langle \hat{a}^\dagger_R (0) \hat{a}_0 (0) \rangle$
and
$G_2(R,t) = g(R,t)- g(R,0)$
with $g(R,t) = \langle \hat{n}_R(t) \hat{n}_0(t) \rangle - \langle \hat{n}_R(t) \rangle \langle \hat{n}_0(t) \rangle$.
Both can be measured in experiments using time-of-flight and fluorescence microscopy imaging, respectively~\cite{cheneau2012,trotzky2012,langen2013,geiger2014}.

All the results presented below are obtained using density-matrix renormalization group simulations
within the time-dependent matrix-product state ($t$-MPS) representation~\cite{schollwock2005,schollwock2011,dolfi2014}.
A careful analysis of the numerical cut-offs (high-filling cut-off and bond dimension) has been systematically performed
to certify the convergence of the results in all the considered cases.
This is particularly critical for quenches in the SF phase where the numerical requirements are most binding~\cite{note:SupplMat}.

\lettersection{Meanfield regime}
\label{sec:SFMF}

\begin{figure}[t!]
\includegraphics[scale=0.22]{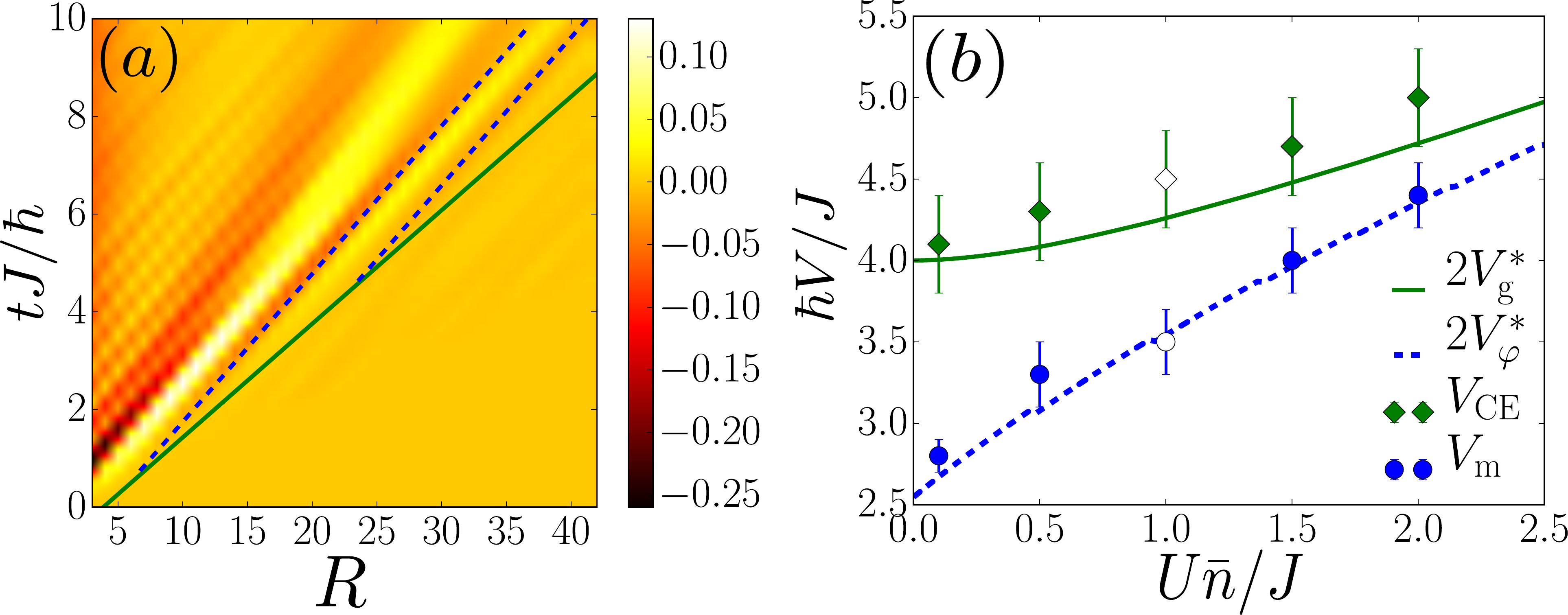}
\caption{\label{fig:velocities_superfluid}
Spreading of correlations in the meanfield regime, see red arrow on Fig.~\ref{fig:phase_diagram}(a).
(a)~$t$-MPS result of $G_2(R,t)$ for a quench to $U/J = 0.1$, together with
ballistic fits to the CE (solid, green line) and minima (dashed, blue lines).
(b)~Velocities of the CE ($\VCE$, green diamonds) and minima ($\Vm$, blue disks), found from the fits, versus the interaction strength,
and comparison to twice the group velocity $2V_\textrm{g}^*$ (solid green line) and twice the phase velocity $2V_{\varphi}^*$ (dashed blue line).
All the quenches are performed with $\qav{n}=5$ from $(U/J)_0 = 0.2$,
except for the points at $U\qav{n}/J=1$ where $U/J=0.2$ and we use a different initial value, $(U/J)_0 = 0.4$ (open points).
}
\end{figure}

We first consider the meanfield regime in the SF phase, where the numerical results can be compared to analytic predictions.
This regime is characterized by a small Lieb-Liniger parameter, $\gamma \equiv U/2J\bar{n} \ll 1$.
Figure~\ref{fig:velocities_superfluid}(a) displays the $t$-MPS result for the $G_2$ correlation function versus distance ($R$) and time ($t$) for a quench from $(U/J)_0 = 0.2$ to $U/J = 0.1$ and $\bar{n}=5$,
\ie\ from $\gamma_0=0.02$ to $\gamma=0.01$ [see red arrow on Fig.~\ref{fig:phase_diagram}(a)].
It clearly shows a spike-like structure, characterized by two different velocities.
On the one hand, a series of parallel maxima and minima move along straight lines corresponding to a constant propagation velocity $\Vm$ (the dashed blue lines show fits to two of these minima).
On the other hand, the various local extrema start at different activation times $t^\star(R)$. The latter are aligned along a straight line with a different slope (solid green line), corresponding a constant velocity $\VCE$. 
The latter defines the correlation edge (CE) beyond which the correlations are suppressed.
Similar results are obtained for all the other quenches in the meanfield regime, as well as for the $G_1$ function~\cite{note:G1mf}.

This twofold structure near the CE is readily interpreted using the quasiparticle picture, which we briefly outline here (for details, see Ref.~\cite{cevolani2018}): 
the $G_1$ and $G_2$ correlation functions are expanded onto the elementary excitations of the system. In the meanfield regime of the BH model,
the latter are Bogoliubov quasiparticles with the quasimomentum $k \in [-\pi,+\pi]$ and the dispersion relation

\begin{equation}
\label{eq:BogoDisp}
E_k \simeq \sqrt{\varepsilon_k \left(\varepsilon_k + 2 \bar{n} U\right)},
\end{equation}

where $\varepsilon_k = 4J \sin^2(k/2)$ is that of the free-particle tight-binding model.
A correlation between two points at a distance $R$ is seeded when two correlated, counter-propagating quasiparticles emanating from the center reach the two points, see Fig.~\ref{fig:phase_diagram}(b). The fastest 
ones are those with the maximum group velocity, $\Vg^\star = \underset{k}{\max} \big( \hbar^{-1}\partial E_k/\partial k \big)$.
It yields the activation time $t^\star(R)=R/2\Vg^\star$ and the CE velocity $\VCE = 2\Vg^\star$,
consistently with the expected Lieb-Robinson bound~\cite{lieb1972,calabrese2006}.
More precisely, the correlation at a distance $R$ and a time $t$ is built as a coherent superposition
of the contributions of the various quasiparticles. In the vicinity of the CE, only the fastest
quasiparticles, \ie\ those with a quasimomentum $k$ close to $k^\star$, contribute. It
creates a sine-like signal at the driving spatial frequency $k^\star$, whose extrema move at
twice the phase velocity $\Vphi(k)=\hbar^{-1} E_k/k$ with $k=k^\star$,
\ie\ $\Vm=2\Vphi^\star$~\cite{cevolani2018}.
The dispersion around $k^\star$ then modulates the sine-like signal by an envelope moving at the CE velocity $\VCE$, see Fig.~\ref{fig:phase_diagram}(c).
This behavior is reminiscent of the propagation of a coherent wave packet in a dispersive medium~\cite{brillouin1960,lighthill1965,born1999}.

To test this picture quantitatively, we have extracted the velocities $\Vm$ and $\VCE$ from the $t$-MPS results for $G_2(R,t)$ by tracking, respectively, the local extrema and the activation time.
The results, displayed on Fig.~\ref{fig:velocities_superfluid}(b), show excellent agreement with the theory,
\ie\ $\VCE \simeq 2\Vg^\star$ and $\Vm \simeq 2\Vphi^\star$ within the fitting errorbars.
This cross-validates the $t$-MPS results in the most-demanding SF, meanfield regime on the one hand and the quasiparticle picture above on the other hand.
Note that the $t$-MPS results are numerically exact and include effects beyond the Bogoliubov approximation,
such as quasiparticle collisions.

\lettersection{Strongly correlated regime at unit filling}

\begin{figure*}[t!]
\includegraphics[width = 2\columnwidth]{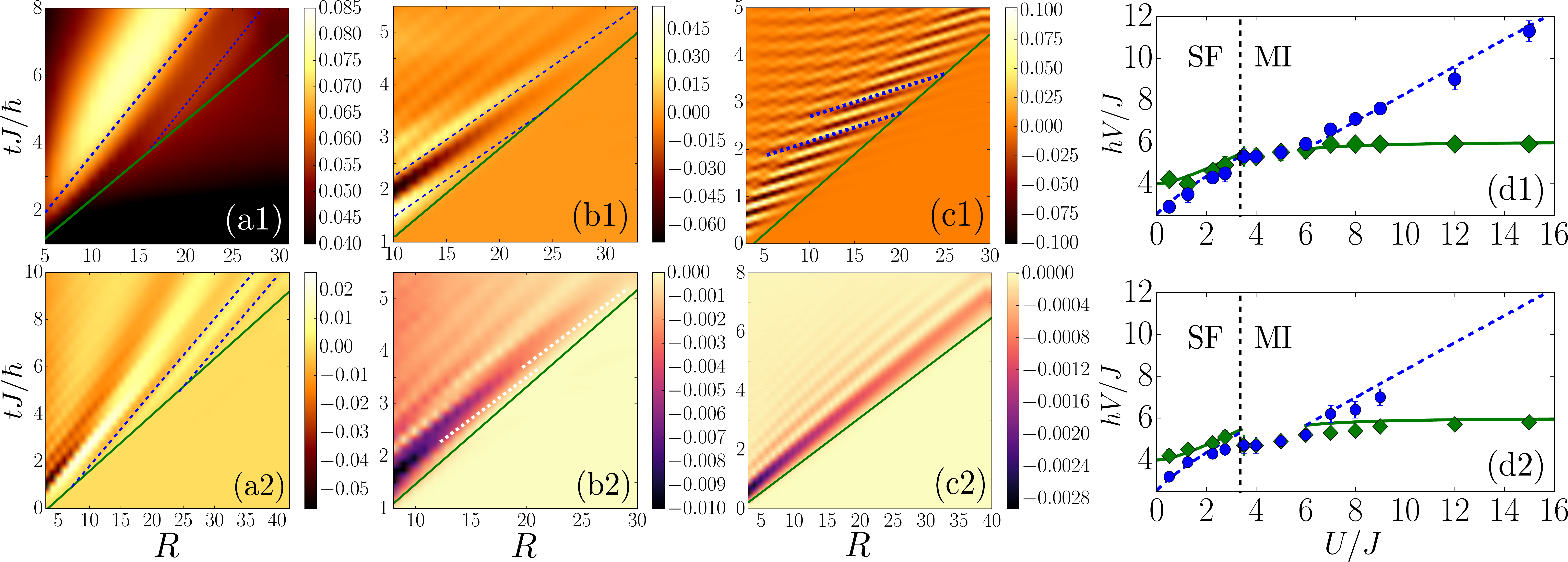}
\caption{\label{fig:numerics}
Spreading of the $G_1$ (upper row) and $G_2$ (lower row) correlations in both the SF and MI phases for $\qav{n}=1$, scanning the after-quench
interaction $U/J$ along the Mott-$U$ transition, see pink dashed line and magenta arrows on Fig.~\ref{fig:phase_diagram}(a):
(a)~SF regime with $U/J = 0.5$;
(b)~MI regime near the critical point with $U/J=8$;
(c)~deep MI regime with $U/J=24$.
The solid green and dashed blue lines correspond to fits to the CE and extrema, respectively.
Note that on panel~(b2), the fits to the maxima are shown as dashed white lines for clarity.
(d)~Spreading velocities $\VCE$ (green diamonds) and $\Vm$ (blue disks), as extracted from fits to the $t$-MPS data,
and comparison to the characteristic velocities $2\Vg^\star$ (solid green lines) and $2\Vphi^\star$ (dashed blue lines), as found from the dispersion relations in the SF~[Eq.~\eqref{eq:BogoDisp}] and MI~[Eq.~\eqref{mott_disp_rel_fermionization}] regimes.
All the quenches are performed from the initial values $(U/J)_0 = 1$ for the SF regime and $(U/J)_0 = \infty$ for the MI regime.
}
\end{figure*}

We now turn to the strongly correlated regime $\gamma \sim 1$, where
the correlation functions cannot be systematically computed.
We first scan the after-quench interaction parameter $U/J$ from the SF to the MI, along the Mott-$U$ transition at unit filling [$\qav{n}=1$, see magenta arrows on Fig.~\ref{fig:phase_diagram}(a)].
Note that each quench is performed in a unique phase: for $U/J<\Ucrit\simeq 3.3$ (SF regime), we use the initial interaction strength $(U/J)_0=1$
while for $U/J>\Ucrit$ (MI regime), we start from $(U/J)_0=\infty$.
Figure~\ref{fig:numerics} shows typical results for the spreading of the $G_1$ (upper row) and $G_2$ (lower row) correlations for quenches
to the SF regime [$U/J=0.5$, Fig.~\ref{fig:numerics}(a)],
and to the MI regime,
both slightly beyond the transition [$U/J=8$, Fig.~\ref{fig:numerics}(b)],
and deep in the MI regime [$U/J=24$, Fig.~\ref{fig:numerics}(c)].
In all cases, at the notable exception of $G_2$ deep in the MI phase [Fig.~\ref{fig:numerics}(c2), see discussion below], we find a twofold spike-like structure.
The velocities $\Vm$ and $\VCE$, extracted as before, are plotted on Fig.~\ref{fig:numerics}(d), showing similar results  for $G_1$ and $G_2$. This is consistent with the prediction that these velocities are characterized by the spectrum,
irrespective of the observable~\cite{cevolani2018}.

In the SF regime, $U/J < \Ucrit$, the results compare very well with the predictions $2\Vphi^\star$ and $2\Vg^\star$ as found from the Bogoliubov dispersion relation~(\ref{eq:BogoDisp}) [see, respectively, the dashed blue and solid green lines on Figs.~\ref{fig:numerics}(d1) and (d2)].
Quite surprizing, the agreement is fair up to the critical point where $\gammacrit \simeq 1.6$, far beyond the validity condition of the Bogoliubov theory ($\gamma \ll 1$).
In fact, when $U/J$ increases from the meanfield regime, the momentum $k^\star$ decreases down to the phonon regime, $k \ll \pi$, and the precise $k$-dependence of the dispersion relation beyond this regime becomes irrelevant. 
Moreover, the physics being dominated by long wavelength excitations, the lattice discretization in Eq.~(\ref{BHm}) may be disregarded and the BH model maps onto the continuous Lieb-Liniger model~\cite{note:SupplMat}.
The latter is integrable by Bethe ansatz (BA)~\cite{lieb1963a,lieb1963b}.
It yields the sound velocity 
$\Vs \simeq 2\qav{n}\sqrt{\gamma}\left(1-\sqrt{\gamma}/4\pi\right)$,
to lowest order in the weak-$\gamma$ expansion.
Up to the critical point, the beyond-meanfield correction, $\sqrt{\gamma}/4\pi$, is less than $10\%$, which explains the good agreement between the numerics and the analytic formula.
At the critical point, the numerical results for $\Vm$ and $\VCE$ are consistent with the exact BA value $2\Vs \simeq 4.6$~\cite{note:Vscrit}.

The spreading velocities $\Vm$ and $\VCE$ are continuous at the Mott-$U$ transition, and do not show any critical behavior. Right beyond the critical point, they are still nearly equal and we can hardly distinguish two features from the numerics up to $U/J \simeq 6$.
Deeper in the MI phase, however, we recover two distinct features and two different velocities.
Contrary to the SF regime, here we find $\Vm>\VCE$.
These results are readily interpreted from the quasiparticle picture.
Deep enough in the MI phase, $U/J \gtrsim 6$, the low-energy excitations are doublon-holon pairs,
characterized by the dispersion relation~\cite{barmettler2012,Ejima2012}

\begin{equation}
2E_k \simeq \sqrt{ \left[ U - 2J(2\bar{n} \! + \! 1)\cos(k)\right]^2 + 16 J^2 \bar{n}(\bar{n} \! + \! 1)\sin^2(k)}.
\label{mott_disp_rel_fermionization}
\end{equation}

The comparison between the spreading velocities $\Vm$ and $\VCE$ fitted from the $t$-MPS results and
the characteristic values $2\Vphi^\star$ and $2\Vg^\star$, found from Eq.~(\ref{mott_disp_rel_fermionization}), 
yields a very good agreement, within less than $5\%$ for $G_1$ and $9\%$ for $G_2$ [see Figs.~\ref{fig:numerics}(d1) and (d2) respectively].
The quantitative agreement between the $t$-MPS results and the theoretical predictions for the $G_1$ correlations persists up to arbitrary values of $U/J$. This validates the quasiparticle analysis also in the strong-coupling regime.

Yet, the $G_2$ correlations behave differently. For intermediate interactions, $6 \lesssim U/J \lesssim 9$, we find a twofold structure consistent with that found for $G_1$.
The signal for $G_2$ blurs when entering deeper in the MI regime, and we are not able to identify two distinct features
for $U/J \gtrsim 9$.
To understand this behavior, one may resort on a strong-coupling ($U \gg J$) expansion of the correlation functions.
In contrast to $G_1$, the $G_2$ function cannot be cast into the generic form analyzed in Ref.~\cite{cevolani2018}.
Instead, combining Jordan-Wigner fermionization and Fermi-Bogoliubov theory~\cite{barmettler2012,note:SupplMat},
one finds $G_2(R,t) \simeq -2|g_2(R,t)|^2$ with 

\begin{equation}
\label{g2_mott}
g_2(R,t) \propto \frac{J}{U}\frac{R}{t} \int_{-\pi}^{+\pi} \frac{\textrm{d}k}{2 \pi} \Big\{\e^{i \left(2E_kt + kR\right)} + \e^{i \left(2E_kt - kR \right)} \Big\}.
\end{equation}

For $U \gg 2(2\bar{n}+1)J$, the doublon-holon pair dispersion relation~(\ref{mott_disp_rel_fermionization}) reduces to
$2E_k \simeq U - 2(2\bar{n}+1)J\cos(k)$. Owing to the square modulus in the formula $G_2(R,t) \simeq -|g_2(R,t)|^2$, we
immediately find that the Mott gap $U$ becomes irrelevant and we are left with the effective dispersion relation $2\tilde{E}_k \simeq
- 2(2\bar{n}+1)J\cos(k)$. On the one hand, the group velocity is not affected and we find the maximum value $2\Vg^\star \simeq
2(2\bar{n}+1)J/\hbar$ at $k^\star \simeq \pi/2$. The value $2\Vg^\star=6J/\hbar$ found for $\qav{n}=1$ is in excellent agreement 
with the  value of $\VCE$ fitted from the $G_2$ function deep in the MI phase, see Fig.~\ref{mott_disp_rel_fermionization}(d2).
On the other hand, the corresponding effective phase velocity vanishes, $2\tildeVphi^\star \simeq 0$. This is consistent with
the disappearance of the spike-like structure observed in the $t$-MPS calculations for $G_2$ deep in the MI phase~\cite{note:ZeroPhaseVelocity}.
In addition, the first-order correction to the leading strong-coupling term, relevant for moderate values of $U/J$, sustains a double structure with $\Vg^\star \neq \Vphi^\star$.
The latter is consistent with the observation of two distinct spreading velocities, $\VCE \neq \Vm$, closer to the
Mott-$U$ transition~\cite{note:SupplMat}.

\lettersection{Strongly interacting superfluid regime}

We finally consider the strongly interacting regime of the SF phase, corresponding to $\gamma \gg 1$ and $\qav{n} \notin\mathbb{N}$. In this regime, the Tomonaga-Luttinger liquid (TLL) theory accurately describes the low-energy physics of the BH model at equilibrium, including the Mott-$\delta$ transition, see for instance Refs.~\cite{cazalilla2011,haller2010,boeris2016}.
The TLL theory considers an effective harmonic fluid, characterized by a single characteristic velocity, namely the sound velocity $\Vs$.

\begin{figure}[t!]
\includegraphics[scale = 0.3]{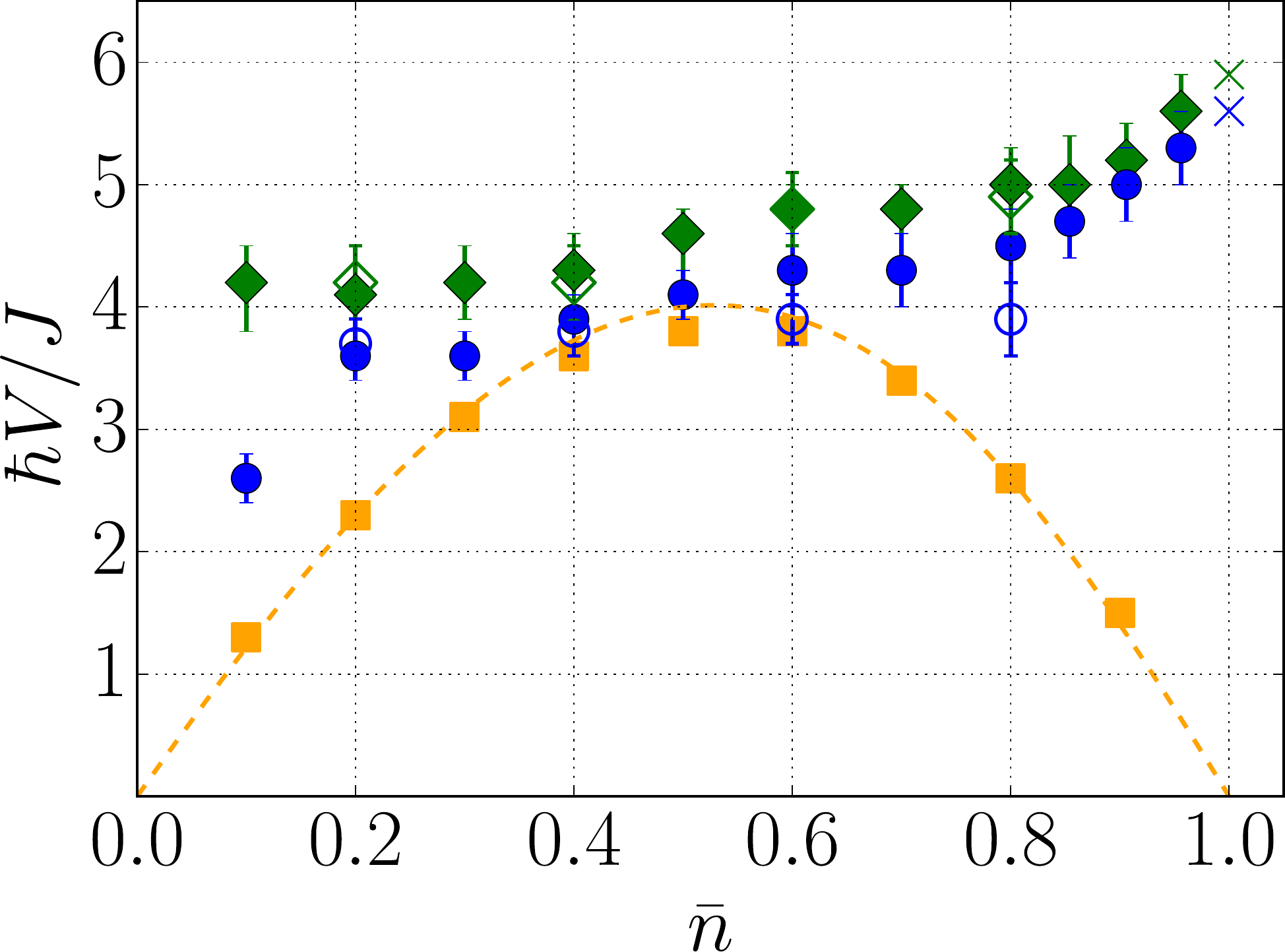}
\caption{ \label{fig:velocities_g2_sir_sf}
Twofold spreading of the $G_2$ correlations in the strongly interacting SF regime for $U/J=50$ and $0<\qav{n}<1$.
Shown are the spreading  velocities $\VCE$ (green diamonds) and $\Vm$ (blue disks) fitted from the $t$-MPS simulations,
together with twice the sound velocity $2\Vs$ of the BH model as found from Bose-Fermi mapping (dashed orange line) and from MPS calculations (orange squares)~\cite{note:VsSISF}.
Filled symbols correspond to the initial interaction parameter $(U/J)_0=1$ and open symbols to $(U/J)_0=40$.
The crosses are linear extrapolations of $\VCE$ and $\Vm$ to the Mott-$\delta$ transition at $\qav{n}=1$.
}
\end{figure}

In contrast, our $t$-MPS simulations in the strongly interacting SF regime clearly show beyond TLL physics.
We have computed the spreading of correlations
for a large value of the after-quench interaction parameter, $U/J=50$, and varying the filling $\qav{n}$
up to the Mott-$\delta$ transition at $\qav{n}=1$ [see pink arrow on Fig.~\ref{fig:phase_diagram}(a)].
The spreading velocities $\VCE$ (green diamonds) and $\Vm$ (blue disks), found from fits to the two-body correlation function $G_2(R,t)$, are shown on Fig.~\ref{fig:velocities_g2_sir_sf}.
They show clear deviations from twice the sound velocity of the BH model in the strongly interacting limit,
$2\Vs \simeq (4J/\hbar)\sin(\pi\qav{n}) \big[1-(8J/U)\cos(\pi\qav{n})\big]$ (orange dotted line and squares)~\cite{note:VsSISF}.
Moreover, the emergence of two different characteristic velocities, $\VCE \neq \Vm$, indicates that the TLL approach is insufficient to describe the spreading of correlations, even upon renormalization of the effective TLL parameters.
Note that the two velocities become nearly equal in the vicinity of the Mott-$\delta$ transition and reach the value $\VCE \simeq \Vm \simeq 6 J/\hbar$. This is consistent with the
disappearance of the twofold structure and the value found for $\VCE$ deep in the MI phase at $\qav{n}=1$, see Fig.~\ref{fig:numerics}(d).

\lettersection{Conclusions}
In summary, working within the case study of the Bose-Hubbard chain and using a numerically-exact many-body approach, we have presented evidence of a universal twofold dynamics in the spreading of correlations. The latter is characterized by two distinct velocities, corresponding to the spreading of local maxima on the one hand and to the CE on the other hand. This has been found in all the phases of the model.
Exceptions appear only in a few cases,
for instance
(i)~for specific observables in specific regimes,
or (ii)~when the two velocities happen to be equal, as found at the Mott critical points for instance.

Our predictions are directly relevant to quench experiments on ultracold Bose gases in optical lattices, where the dynamics of one-body and two-body correlation functions can be observed on space and time scales comparable to our simulations~\cite{lewenstein2007, bloch2008, cheneau2012, trotzky2012, NaturePhysicsInsight2012bloch}.
Importantly, while in most experiments and numerics the CE is infered from the behavior of the correlation maxima, our results show that the two must be distinguished. This is expected to be a general feature of short-range systems and should be relevant to models other than the sole BH model.

Moreover, our study may be extended to long-range systems, such as spin models as realized in trapped-ion experiments~\cite{jurcevic2014,richerme2014}.
While the notions of a maximum group velocity and phase velocity may break down in such systems, the meanfield theory also predicts a twofold dynamics~\cite{cevolani2018}. In this case, it is characterized by the coexistence of super-ballistic and sub-ballistic signals. The results of the present paper suggest that the twofold structure of the correlation function may survive in strongly correlated regimes also for long-range systems.
The demonstration of this effect would shed light on the still debated scaling of the light cone in long-range systems.

\acknowledgments
This research was supported by the
European Commission FET-Proactive QUIC (H2020 grant No.~641122).
The numerical calculations were performed using HPC resources 
from CPHT and GENCI-CCRT/CINES (Grant No.~c2017056853),
and make use of the ALPS library~\cite{dolfi2014}.
We are grateful to  the CPHT computer team for valuable support.


\begin{thebibliography}{10}

\providecommand*{\bibinfo}[2]{#2}
\providecommand*{\eprint}[1]{#1}
\providecommand*{\url}[1]{#1}
\bibitem{polkovnikov2011}
\bibinfo{author}{A.~Polkovnikov}, \bibinfo{author}{K.~Sengupta},
  \bibinfo{author}{A.~Silva}, and \bibinfo{author}{M.~Vengalattore},
  \bibinfo{title}{\emph{\textit{Colloquium}: Nonequilibrium dynamics of closed
  interacting quantum systems}}, \bibinfo{journal}{\Jrmp}
  \bibinfo{volume}{\textbf{83}}, \bibinfo{pages}{863} (\bibinfo{date}{2011}).
\bibitem{eisert2015}
\bibinfo{author}{J.~Eisert}, \bibinfo{author}{M.~Friesdorf}, and
  \bibinfo{author}{C.~Gogolin}, \bibinfo{title}{\emph{Quantum many-body systems
  out of equilibrium}}, \bibinfo{journal}{\Jnatphys}
  \bibinfo{volume}{\textbf{11}}, \bibinfo{pages}{124} (\bibinfo{date}{2015}).
\bibitem{pekola2015}
\bibinfo{author}{J.~P. Pekola}, \bibinfo{title}{\emph{Towards quantum
  thermodynamics in electronic circuits}}, \bibinfo{journal}{\Jnatphys}
  \bibinfo{volume}{\textbf{11}}, \bibinfo{pages}{118} (\bibinfo{date}{2015}).
\bibitem{langen2015}
\bibinfo{author}{T.~{Langen}}, \bibinfo{author}{R.~{Geiger}}, and
  \bibinfo{author}{J.~{Schmiedmayer}}, \bibinfo{title}{\emph{Ultracold atoms
  out of equilibrium}}, \bibinfo{journal}{\JAnnualRevCondMat}
  \bibinfo{volume}{\textbf{6}}, \bibinfo{pages}{201} (\bibinfo{date}{2015}).
\bibitem{lewenstein2007}
\bibinfo{author}{M.~Lewenstein}, \bibinfo{author}{A.~Sanpera},
  \bibinfo{author}{V.~Ahufinger}, \bibinfo{author}{B.~Damski},
  \bibinfo{author}{A.~Sen}, and \bibinfo{author}{U.~Sen},
  \bibinfo{title}{\emph{Ultracold atomic gases in optical lattices: {M}imicking
  condensed matter physics and beyond}}, \bibinfo{journal}{\Jadvphys}
  \bibinfo{volume}{\textbf{56}}, \bibinfo{pages}{243} (\bibinfo{date}{2007}).
\bibitem{bloch2008}
\bibinfo{author}{I.~Bloch}, \bibinfo{author}{J.~Dalibard}, and
  \bibinfo{author}{W.~Zwerger}, \bibinfo{title}{\emph{Many-body physics with
  ultracold gases}}, \bibinfo{journal}{\Jrmp} \bibinfo{volume}{\textbf{80}},
  \bibinfo{pages}{885} (\bibinfo{date}{2008}).
\bibitem{NaturePhysicsInsight2012bloch}
\bibinfo{author}{I.~Bloch}, \bibinfo{author}{J.~Dalibard}, and
  \bibinfo{author}{S.~Nascimb\`ene}, \bibinfo{title}{\emph{Quantum simulations
  with ultracold quantum gases}}, \bibinfo{journal}{\Jnatphys}
  \bibinfo{volume}{\textbf{8}}, \bibinfo{pages}{267} (\bibinfo{date}{2012}).
\bibitem{NaturePhysicsInsight2012blatt}
\bibinfo{author}{R.~Blatt} and \bibinfo{author}{C.~F. Roos},
  \bibinfo{title}{\emph{Quantum simulations with trapped ions}},
  \bibinfo{journal}{\Jnatphys} \bibinfo{volume}{\textbf{8}},
  \bibinfo{pages}{277} (\bibinfo{date}{2012}).
\bibitem{NaturePhysicsInsight2012aspuru-guzik}
\bibinfo{author}{A.~Aspuru-Guzik} and \bibinfo{author}{P.~Walther},
  \bibinfo{title}{\emph{Photonic quantum simulators}},
  \bibinfo{journal}{\Jnatphys} \bibinfo{volume}{\textbf{8}},
  \bibinfo{pages}{285} (\bibinfo{date}{2012}).
\bibitem{NaturePhysicsInsight2012houck}
\bibinfo{author}{A.~A. Houck}, \bibinfo{author}{H.~E. Tureci}, and
  \bibinfo{author}{J.~Koch}, \bibinfo{title}{\emph{On-chip quantum simulation
  with superconducting circuits}}, \bibinfo{journal}{\Jnatphys}
  \bibinfo{volume}{\textbf{8}}, \bibinfo{pages}{292} (\bibinfo{date}{2012}).
\bibitem{lieb1972}
\bibinfo{author}{E.~H. Lieb} and \bibinfo{author}{D.~W. Robinson},
  \bibinfo{title}{\emph{The finite group velocity of quantum spin systems}},
  \bibinfo{journal}{\JCommMathPhys} \bibinfo{volume}{\textbf{28}},
  \bibinfo{pages}{251} (\bibinfo{date}{1972}).
\bibitem{bravyi2006}
\bibinfo{author}{S.~Bravyi}, \bibinfo{author}{M.~B. Hastings}, and
  \bibinfo{author}{F.~Verstraete}, \bibinfo{title}{\emph{{L}ieb-{R}obinson
  bounds and the generation of correlations and topological quantum order}},
  \bibinfo{journal}{\Jprl} \bibinfo{volume}{\textbf{97}},
  \bibinfo{pages}{050401} (\bibinfo{date}{2006}).
\bibitem{hastings2006}
\bibinfo{author}{M.~B. Hastings} and \bibinfo{author}{T.~Koma},
  \bibinfo{title}{\emph{Spectral gap and exponential decay of correlations}},
  \bibinfo{journal}{\JCommMathPhys} \bibinfo{volume}{\textbf{265}},
  \bibinfo{pages}{781} (\bibinfo{date}{2006}).
\bibitem{barmettler2012}
\bibinfo{author}{P.~Barmettler}, \bibinfo{author}{D.~Poletti},
  \bibinfo{author}{M.~Cheneau}, and \bibinfo{author}{C.~Kollath},
  \bibinfo{title}{\emph{Propagation front of correlations in an interacting
  {B}ose gas}}, \bibinfo{journal}{\Jpra} \bibinfo{volume}{\textbf{85}},
  \bibinfo{pages}{053625} (\bibinfo{date}{2012}).
\bibitem{cheneau2012}
\bibinfo{author}{M.~Cheneau}, \bibinfo{author}{P.~Barmettler},
  \bibinfo{author}{D.~Poletti}, \bibinfo{author}{M.~Endres},
  \bibinfo{author}{P.~Schauss}, \bibinfo{author}{T.~Fukuhara},
  \bibinfo{author}{C.~Gross}, \bibinfo{author}{I.~Bloch},
  \bibinfo{author}{C.~Kollath}, and \bibinfo{author}{S.~Kuhr},
  \bibinfo{title}{\emph{Light-cone-like spreading of correlations in a quantum
  many-body system}}, \bibinfo{journal}{\Jnature}
  \bibinfo{volume}{\textbf{481}}, \bibinfo{pages}{484} (\bibinfo{date}{2012}).
\bibitem{carleo2014}
\bibinfo{author}{G.~Carleo}, \bibinfo{author}{F.~Becca},
  \bibinfo{author}{L.~Sanchez-Palencia}, \bibinfo{author}{S.~Sorella}, and
  \bibinfo{author}{M.~Fabrizio}, \bibinfo{title}{\emph{Light-cone effect and
  supersonic correlations in one- and two-dimensional bosonic superfluids}},
  \bibinfo{journal}{\Jpra} \bibinfo{volume}{\textbf{89}},
  \bibinfo{pages}{031602(R)} (\bibinfo{date}{2014}).
\bibitem{manmana2009}
\bibinfo{author}{S.~R. Manmana}, \bibinfo{author}{S.~Wessel},
  \bibinfo{author}{R.~M. Noack}, and \bibinfo{author}{A.~Muramatsu},
  \bibinfo{title}{\emph{Time evolution of correlations in strongly interacting
  fermions after a quantum quench}}, \bibinfo{journal}{\Jprb}
  \bibinfo{volume}{\textbf{79}}, \bibinfo{pages}{155104}
  (\bibinfo{date}{2009}).
\bibitem{jurcevic2014}
\bibinfo{author}{P.~Jurcevic}, \bibinfo{author}{B.~P. Lanyon},
  \bibinfo{author}{P.~Hauke}, \bibinfo{author}{C.~Hempel},
  \bibinfo{author}{P.~Zoller}, \bibinfo{author}{R.~Blatt}, and
  \bibinfo{author}{C.~F. Roos}, \bibinfo{title}{\emph{Quasiparticle engineering
  and entanglement propagation in a quantum many-body system}},
  \bibinfo{journal}{\Jnature} \bibinfo{volume}{\textbf{511}},
  \bibinfo{pages}{202} (\bibinfo{date}{2014}).
\bibitem{richerme2014}
\bibinfo{author}{P.~Richerme}, \bibinfo{author}{Z.-X. Gong},
  \bibinfo{author}{A.~Lee}, \bibinfo{author}{C.~Senko},
  \bibinfo{author}{J.~Smith}, \bibinfo{author}{M.~Foss-Feig},
  \bibinfo{author}{S.~Michalakis}, \bibinfo{author}{A.~V. Gorshkov}, and
  \bibinfo{author}{C.~Monroe}, \bibinfo{title}{\emph{Non-local propagation of
  correlations in quantum systems with long-range interactions}},
  \bibinfo{journal}{\Jnature} \bibinfo{volume}{\textbf{511}},
  \bibinfo{pages}{198} (\bibinfo{date}{2014}).
\bibitem{hauke2013}
\bibinfo{author}{P.~Hauke} and \bibinfo{author}{L.~Tagliacozzo},
  \bibinfo{title}{\emph{Spread of correlations in long-range interacting
  quantum systems}}, \bibinfo{journal}{\Jprl} \bibinfo{volume}{\textbf{111}},
  \bibinfo{pages}{207202} (\bibinfo{date}{2013}).
\bibitem{eisert2013}
\bibinfo{author}{J.~Eisert}, \bibinfo{author}{M.~van~den Worm},
  \bibinfo{author}{S.~R. Manmana}, and \bibinfo{author}{M.~Kastner},
  \bibinfo{title}{\emph{Breakdown of quasilocality in long-range quantum
  lattice models}}, \bibinfo{journal}{\Jprl} \bibinfo{volume}{\textbf{111}},
  \bibinfo{pages}{260401} (\bibinfo{date}{2013}).
\bibitem{cevolani2015}
\bibinfo{author}{L.~Cevolani}, \bibinfo{author}{G.~Carleo}, and
  \bibinfo{author}{L.~Sanchez-Palencia}, \bibinfo{title}{\emph{Protected
  quasi-locality in quantum systems with long-range interactions}},
  \bibinfo{journal}{\Jpra} \bibinfo{volume}{\textbf{92}},
  \bibinfo{pages}{041603(R)} (\bibinfo{date}{2015}).
\bibitem{schachenmayer2015b}
\bibinfo{author}{J.~Schachenmayer}, \bibinfo{author}{A.~Pikovski}, and
  \bibinfo{author}{A.~M. Rey}, \bibinfo{title}{\emph{Dynamics of correlations
  in two-dimensional quantum spin models with long-range interactions: {A}
  phase-space {M}onte-{C}arlo study}}, \bibinfo{journal}{\Jnjp}
  \bibinfo{volume}{\textbf{17}}(6), \bibinfo{pages}{065009}
  (\bibinfo{date}{2015}).
\bibitem{buyskikh2016}
\bibinfo{author}{A.~S. Buyskikh}, \bibinfo{author}{M.~Fagotti},
  \bibinfo{author}{J.~Schachenmayer}, \bibinfo{author}{F.~Essler}, and
  \bibinfo{author}{A.~J. Daley}, \bibinfo{title}{\emph{Entanglement growth and
  correlation spreading with variable-range interactions in spin and fermionic
  tunneling models}}, \bibinfo{journal}{\Jpra} \bibinfo{volume}{\textbf{93}},
  \bibinfo{pages}{053620} (\bibinfo{date}{2016}).
\bibitem{cevolani2016}
\bibinfo{author}{L.~Cevolani}, \bibinfo{author}{G.~Carleo}, and
  \bibinfo{author}{L.~Sanchez-Palencia}, \bibinfo{title}{\emph{Spreading of
  correlations in exactly solvable quantum models with long-range interactions
  in arbitrary dimensions}}, \bibinfo{journal}{\Jnjp}
  \bibinfo{volume}{\textbf{18}}, \bibinfo{pages}{093002}
  (\bibinfo{date}{2016}).
\bibitem{frerot2018}
\bibinfo{author}{I.~Fr\'erot}, \bibinfo{author}{P.~Naldesi}, and
  \bibinfo{author}{T.~Roscilde}, \bibinfo{title}{\emph{Multispeed
  prethermalization in quantum spin models with power-law decaying
  interactions}}, \bibinfo{journal}{\Jprl} \bibinfo{volume}{\textbf{120}},
  \bibinfo{pages}{050401} (\bibinfo{date}{2018}).
\bibitem{cevolani2018}
\bibinfo{author}{L.~Cevolani}, \bibinfo{author}{J.~Despres},
  \bibinfo{author}{G.~Carleo}, \bibinfo{author}{L.~Tagliacozzo}, and
  \bibinfo{author}{L.~Sanchez-Palencia}, \bibinfo{title}{\emph{Universal
  scaling laws for correlation spreading in quantum systems with short- and
  long-range interactions}}, \bibinfo{journal}{\Jprb}
  \bibinfo{volume}{\textbf{98}}, \bibinfo{pages}{024302}
  (\bibinfo{date}{2018}).
\bibitem{foss-feig2015}
\bibinfo{author}{M.~Foss-Feig}, \bibinfo{author}{Z.-X. Gong},
  \bibinfo{author}{C.~W. Clark}, and \bibinfo{author}{A.~V. Gorshkov},
  \bibinfo{title}{\emph{Nearly linear light cones in long-range interacting
  quantum systems}}, \bibinfo{journal}{\Jprl} \bibinfo{volume}{\textbf{114}},
  \bibinfo{pages}{157201} (\bibinfo{date}{2015}).
\bibitem{sachdev2001}
\bibinfo{author}{S.~Sachdev}, \bibinfo{title}{\emph{Quantum Phase Transitions}}
  (\bibinfo{publisher}{Cambridge University Press, Cambridge, UK},
  \bibinfo{year}{2001}).
\bibitem{cazalilla2011}
\bibinfo{author}{M.~A. Cazalilla}, \bibinfo{author}{R.~Citro},
  \bibinfo{author}{T.~Giamarchi}, \bibinfo{author}{E.~Orignac}, and
  \bibinfo{author}{M.~Rigol}, \bibinfo{title}{\emph{One dimensional bosons:
  From condensed matter systems to ultracold gases}}, \bibinfo{journal}{\Jrmp}
  \bibinfo{volume}{\textbf{83}}, \bibinfo{pages}{1405} (\bibinfo{date}{2011}).
\bibitem{kuhner2000}
\bibinfo{author}{T.~D. K\"uhner}, \bibinfo{author}{S.~R. White}, and
  \bibinfo{author}{H.~Monien}, \bibinfo{title}{\emph{One-dimensional
  {B}ose-{H}ubbard model with nearest-neighbor interaction}},
  \bibinfo{journal}{\prb} \bibinfo{volume}{\textbf{61}}, \bibinfo{pages}{12474}
  (\bibinfo{date}{2000}).
\bibitem{kashurnikov1996exact}
\bibinfo{author}{V.~Kashurnikov} and \bibinfo{author}{B.~Svistunov},
  \bibinfo{title}{\emph{Exact diagonalization plus renormalization-group
  theory: Accurate method for a one-dimensional superfluid-insulator-transition
  study}}, \bibinfo{journal}{\Jprb} \bibinfo{volume}{\textbf{53}},
  \bibinfo{pages}{11776} (\bibinfo{date}{1996}).
\bibitem{ejima2011}
\bibinfo{author}{S.~Ejima}, \bibinfo{author}{H.~Fehske}, and
  \bibinfo{author}{F.~Gebhard}, \bibinfo{title}{\emph{Dynamic properties of the
  one-dimensional bose-hubbard model}}, \bibinfo{journal}{\Jepl}
  \bibinfo{volume}{\textbf{93}}, \bibinfo{pages}{30002} (\bibinfo{date}{2011}).
\bibitem{rombouts2006}
\bibinfo{author}{S.~Rombouts}, \bibinfo{author}{K.~Van~Houcke}, and
  \bibinfo{author}{L.~Pollet}, \bibinfo{title}{\emph{Loop updates for quantum
  {M}onte {C}arlo simulations in the canonical ensemble}},
  \bibinfo{journal}{\Jprl} \bibinfo{volume}{\textbf{96}},
  \bibinfo{pages}{180603} (\bibinfo{date}{2006}).
\bibitem{calabrese2006}
\bibinfo{author}{P.~Calabrese} and \bibinfo{author}{J.~Cardy},
  \bibinfo{title}{\emph{Time dependence of correlation functions following a
  quantum quench}}, \bibinfo{journal}{\Jprl} \bibinfo{volume}{\textbf{96}},
  \bibinfo{pages}{136801} (\bibinfo{date}{2006}).
\bibitem{kollath2007}
\bibinfo{author}{C.~Kollath}, \bibinfo{author}{A.~M. L\"auchli}, and
  \bibinfo{author}{E.~Altman}, \bibinfo{title}{\emph{Quench dynamics and
  nonequilibrium phase diagram of the {B}ose-{H}ubbard model}},
  \bibinfo{journal}{\Jprl} \bibinfo{volume}{\textbf{98}},
  \bibinfo{pages}{180601} (\bibinfo{date}{2007}).
\bibitem{moeckel2008}
\bibinfo{author}{M.~Moeckel} and \bibinfo{author}{S.~Kehrein},
  \bibinfo{title}{\emph{Interaction quench in the {H}ubbard model}},
  \bibinfo{journal}{\Jprl} \bibinfo{volume}{\textbf{100}},
  \bibinfo{pages}{175702} (\bibinfo{date}{2008}).
\bibitem{roux2010}
\bibinfo{author}{G.~Roux}, \bibinfo{title}{\emph{Finite-size effects in global
  quantum quenches: Examples from free bosons in an harmonic trap and the
  one-dimensional bose-hubbard model}}, \bibinfo{journal}{\Jpra}
  \bibinfo{volume}{\textbf{81}}, \bibinfo{pages}{053604}
  (\bibinfo{date}{2010}).
\bibitem{navez2010}
\bibinfo{author}{P.~Navez} and \bibinfo{author}{R.~Sch\"utzhold},
  \bibinfo{title}{\emph{Emergence of coherence in the {M}ott
  insulator-superfluid quench of the {B}ose-{H}ubbard model}},
  \bibinfo{journal}{\Jpra} \bibinfo{volume}{\textbf{82}},
  \bibinfo{pages}{063603} (\bibinfo{date}{2010}).
\bibitem{krutitsky2014}
\bibinfo{author}{K.~V. Krutitsky}, \bibinfo{author}{P.~Navez},
  \bibinfo{author}{F.~Queisser}, and \bibinfo{author}{R.~Sch{\"u}tzhold},
  \bibinfo{title}{\emph{Propagation of quantum correlations after a quench in
  the {M}ott insulator regime of the {B}ose-{H}ubbard model}},
  \bibinfo{journal}{EPJ Quantum Technology} \bibinfo{volume}{\textbf{1}},
  \bibinfo{pages}{12} (\bibinfo{date}{2014}).
\bibitem{greiner2002b}
\bibinfo{author}{M.~Greiner}, \bibinfo{author}{O.~Mandel}, ,
  \bibinfo{author}{T.~W. H\"ansch}, and \bibinfo{author}{I.~Bloch},
  \bibinfo{title}{\emph{Collapse and revival of the matter wave field of a
  {B}ose-{E}instein condensate}}, \bibinfo{journal}{\Jnature}
  \bibinfo{volume}{\textbf{419}}, \bibinfo{pages}{51} (\bibinfo{date}{2002}).
\bibitem{langen2013}
\bibinfo{author}{T.~Langen}, \bibinfo{author}{R.~Geiger},
  \bibinfo{author}{M.~Kuhnert}, \bibinfo{author}{B.~Rauer}, and
  \bibinfo{author}{J.~Schmiedmayer}, \bibinfo{title}{\emph{Local emergence of
  thermal correlations in an isolated quantum many-body system}},
  \bibinfo{journal}{\Jnatphys} \bibinfo{volume}{\textbf{9}},
  \bibinfo{pages}{640} (\bibinfo{date}{2013}).
\bibitem{geiger2014}
\bibinfo{author}{R.~Geiger}, \bibinfo{author}{T.~Langen},
  \bibinfo{author}{I.~E. Mazets}, and \bibinfo{author}{J.~Schmiedmayer},
  \bibinfo{title}{\emph{Local relaxation and light-cone-like propagation of
  correlations in a trapped one-dimensional {B}ose gas}},
  \bibinfo{journal}{\Jnjp} \bibinfo{volume}{\textbf{16}},
  \bibinfo{pages}{053034} (\bibinfo{date}{2014}).
\bibitem{trotzky2012}
\bibinfo{author}{S.~Trotzky}, \bibinfo{author}{Y.-A. Chen},
  \bibinfo{author}{A.~Flesch}, \bibinfo{author}{I.~P. McCulloch},
  \bibinfo{author}{U.~Schollw\"ock}, \bibinfo{author}{J.~Eisert}, and
  \bibinfo{author}{I.~Bloch}, \bibinfo{title}{\emph{Probing the relaxation
  towards equilibrium in an isolated strongly correlated one-dimensional {B}ose
  gas}}, \bibinfo{journal}{\Jnatphys} \bibinfo{volume}{\textbf{8}},
  \bibinfo{pages}{325} (\bibinfo{date}{2012}).
\bibitem{schollwock2005}
\bibinfo{author}{U.~Schollw\"ock}, \bibinfo{title}{\emph{The density-matrix
  renormalization group}}, \bibinfo{journal}{\Jrmp}
  \bibinfo{volume}{\textbf{77}}, \bibinfo{pages}{259} (\bibinfo{date}{2005}).
\bibitem{schollwock2011}
\bibinfo{author}{U.~Schollw\"ock}, \bibinfo{title}{\emph{The density-matrix
  renormalization group in the age of matrix product states}},
  \bibinfo{journal}{\Jannphys} \bibinfo{volume}{\textbf{326}},
  \bibinfo{pages}{96} (\bibinfo{date}{2011}).
\bibitem{dolfi2014}
\bibinfo{author}{M.~Dolfi}, \bibinfo{author}{B.~Bauer},
  \bibinfo{author}{S.~Keller}, \bibinfo{author}{A.~Kosenkov},
  \bibinfo{author}{T.~Ewart}, \bibinfo{author}{A.~Kantian},
  \bibinfo{author}{T.~Giamarchi}, and \bibinfo{author}{M.~Troyer},
  \bibinfo{title}{\emph{Matrix product state applications for the {ALPS}
  project}}, \bibinfo{journal}{Comput. Phys. Commun.}
  \bibinfo{volume}{\textbf{185}}, \bibinfo{pages}{3430 }
  (\bibinfo{date}{2014}).
\bibitem{note:SupplMat}
{For further details, see Supplemental Material. It contains information about
  the $t$-MPS calculations, the spreading of the one-body correlations ($G_1$)
  in the meanfield regime and two-body correlations ($G_2$) in the deep MI
  phase, as well as the mapping onto the Lieb-Liniger model.}
\bibitem{note:G1mf}
{Note that the signal for $G_1$ is, however, less sharp than for $G_2$. This
  may be attributed to the long-range correlations present in the initial
  state, which partially blur the CE~\cite{note:SupplMat} }.
\bibitem{brillouin1960}
\bibinfo{author}{L.~Brillouin}, \bibinfo{title}{\emph{Wave propagation and
  group velocity}} (\bibinfo{publisher}{Academic Press}, \bibinfo{year}{1960}).
\bibitem{lighthill1965}
\bibinfo{author}{M.~Lighthill}, \bibinfo{title}{\emph{Group velocity}}
  (\bibinfo{publisher}{Oxford University Press}, \bibinfo{year}{1965}).
\bibitem{born1999}
\bibinfo{author}{M.~Born} and \bibinfo{author}{E.~Wolf},
  \bibinfo{title}{\emph{Principles of Optics, 7-th edition}}
  (\bibinfo{publisher}{Cambridge University Press, Cambridge},
  \bibinfo{year}{1999}).
\bibitem{lieb1963a}
\bibinfo{author}{E.~H. Lieb} and \bibinfo{author}{W.~Liniger},
  \bibinfo{title}{\emph{Exact analysis of an interacting {B}ose gas. {I}. {T}he
  general solution and the ground state}}, \bibinfo{journal}{\Jpr}
  \bibinfo{volume}{\textbf{130}}, \bibinfo{pages}{1605} (\bibinfo{date}{1963}).
\bibitem{lieb1963b}
\bibinfo{author}{E.~H. Lieb}, \bibinfo{title}{\emph{Exact analysis of an
  interacting {B}ose gas. {II}. {T}he excitation spectrum}},
  \bibinfo{journal}{\Jpr} \bibinfo{volume}{\textbf{130}}, \bibinfo{pages}{1616}
  (\bibinfo{date}{1963}).
\bibitem{note:Vscrit}
{Close to the Mott-$U$ critical point at $U/J=3.5$, we find $\Vm \simeq \VCE
  \simeq 4.7$ ($5.3$) for the $G_2$ ($G_1$) correlation function, which agrees
  with the value of $2\Vs$ within $2\%$ ($13\%$).}
\bibitem{Ejima2012}
\bibinfo{author}{S.~Ejima}, \bibinfo{author}{H.~Fehske},
  \bibinfo{author}{F.~Gebhard}, \bibinfo{author}{K.~zu~M{\"u}nster},
  \bibinfo{author}{M.~Knap}, \bibinfo{author}{E.~Arrigoni}, and
  \bibinfo{author}{W.~von~der Linden}, \bibinfo{title}{\emph{Characterization
  of {M}ott-insulating and superfluid phases in the one-dimensional
  {B}ose-{H}ubbard model}}, \bibinfo{journal}{\Jpra}
  \bibinfo{volume}{\textbf{85}}, \bibinfo{pages}{053644}
  (\bibinfo{date}{2012}).
\bibitem{note:ZeroPhaseVelocity}
{ More precisely, we find that in the vicinity of the CE both the real and
  imaginary parts of $g_2$ display a series of static local maxima,
  consistently with $2\tildeVphi^\star \simeq 0$. These local maxima are
  shifted by half a period and cancel each other when combined for constructing
  $G_2$~\cite{note:SupplMat} }.
\bibitem{haller2010}
\bibinfo{author}{E.~Haller}, \bibinfo{author}{R.~Hart}, \bibinfo{author}{M.~J.
  Mark}, \bibinfo{author}{J.~G. Danzl}, \bibinfo{author}{L.~Reichs\"ollner},
  \bibinfo{author}{M.~Gustavsson}, \bibinfo{author}{M.~Dalmonte},
  \bibinfo{author}{G.~Pupillo}, and \bibinfo{author}{H.-C. N\"agerl},
  \bibinfo{title}{\emph{Pinning quantum phase transition for a {L}uttinger
  liquid of strongly interacting bosons}}, \bibinfo{journal}{\Jnature}
  \bibinfo{volume}{\textbf{466}}, \bibinfo{pages}{597} (\bibinfo{date}{2010}).
\bibitem{boeris2016}
\bibinfo{author}{G.~Bo\'eris}, \bibinfo{author}{L.~Gori},
  \bibinfo{author}{M.~D. Hoogerland}, \bibinfo{author}{A.~Kumar},
  \bibinfo{author}{E.~Lucioni}, \bibinfo{author}{L.~Tanzi},
  \bibinfo{author}{M.~Inguscio}, \bibinfo{author}{T.~Giamarchi},
  \bibinfo{author}{C.~D'Errico}, \bibinfo{author}{G.~Carleo}, \emph{et~al.},
  \bibinfo{title}{\emph{{M}ott transition for strongly interacting
  one-dimensional bosons in a shallow periodic potential}},
  \bibinfo{journal}{\Jpra} \bibinfo{volume}{\textbf{93}},
  \bibinfo{pages}{011601(R)} (\bibinfo{date}{2016}).
\bibitem{note:VsSISF}
{ The sound velocity $\Vs$ has been computed by mapping the BH model to an
  equivalent spinless Fermi model~\cite{cazalilla2003,cazalilla2004} (dotted
  orange line) and, independently, from the energy of the first excited state
  in exact MPS calculations (see orange squares), showing excellent agreement.}
\bibitem{cazalilla2003}
\bibinfo{author}{M.~A. Cazalilla}, \bibinfo{title}{\emph{One-dimensional
  optical lattices and impenetrable bosons}}, \bibinfo{journal}{\Jpra}
  \bibinfo{volume}{\textbf{67}}, \bibinfo{pages}{053606}
  (\bibinfo{date}{2003}).
\bibitem{cazalilla2004}
\bibinfo{author}{M.~A. Cazalilla}, \bibinfo{title}{\emph{Differences between
  the {T}onks regimes in the continuum and on the lattice}},
  \bibinfo{journal}{\Jpra} \bibinfo{volume}{\textbf{70}},
  \bibinfo{pages}{041604} (\bibinfo{date}{2004}).
\bibitem{wolf2008}
\bibinfo{author}{M.~M. Wolf}, \bibinfo{author}{F.~Verstraete},
  \bibinfo{author}{M.~B. Hastings}, and \bibinfo{author}{J.~I. Cirac},
  \bibinfo{title}{\emph{Area laws in quantum systems: Mutual information and
  correlations}}, \bibinfo{journal}{\Jprl} \bibinfo{volume}{\textbf{100}},
  \bibinfo{pages}{070502} (\bibinfo{date}{2008}).
\bibitem{eisert2010}
\bibinfo{author}{J.~Eisert}, \bibinfo{author}{M.~Cramer}, and
  \bibinfo{author}{M.~B. Plenio}, \bibinfo{title}{\emph{\textit{Colloquium}:
  Area laws for the entanglement entropy}}, \bibinfo{journal}{\Jrmp}
  \bibinfo{volume}{\textbf{82}}, \bibinfo{pages}{277} (\bibinfo{date}{2010}).
\bibitem{kashurnikov1998zero}
\bibinfo{author}{V.~A. Kashurnikov}, \bibinfo{author}{A.~V. Krasavin}, and
  \bibinfo{author}{B.~V. Svistunov}, \bibinfo{title}{\emph{Zero-point phase
  transitions in the one-dimensional truncated bosonic {H}ubbard model and its
  spin-1 analog}}, \bibinfo{journal}{\Jprb} \bibinfo{volume}{\textbf{58}},
  \bibinfo{pages}{1826} (\bibinfo{date}{1998}).

\end{thebibliography}

  
 \renewcommand{\theequation}{S\arabic{equation}}
 \setcounter{equation}{0}
 \renewcommand{\thefigure}{S\arabic{figure}}
 \setcounter{figure}{0}
 \renewcommand{\thesection}{S\arabic{section}}
 \setcounter{section}{0}
 \onecolumngrid  
 
 \newpage

 {\center \bf \large Supplemental Material for \\}
 {\center \bf \large \ttitle \\ \vspace*{1.cm}
 }
 
In this supplemental material, we give more details about several points discussed in the main paper.
In Sec.~\ref{tMPS}, we discuss the time-dependent matrix-product state ($t$-MPS) simulations and the choice of the numerical
parameters to ensure the convergence of the numerical results in all the regimes of the Bose-Hubbard model considered in the main paper.
In Sec.~\ref{g1_mf}, we present $t$-MPS results for the spreading of the one-body correlation function $G_1$ in the meanfield superfluid (SF) regime.
Section~\ref{mapping} briefly outlines the mapping from the 1D Bose-Hubbard model to the Lieb-Liniger model and gives the 
correspondance of the parameters.
Finally, in Sec.~\ref{spa} we discuss the strong-coupling expansion of the correlation function $G_2$ for unit filling, $\qav{n}=1$, and discuss the 
suppression of its twofold structure deep in the Mott insulator (MI) phase.

\section{Time-dependent matrix-product state simulations}
\label{tMPS}

The numerical results reported in the main paper are all obtained using the time-dependent density-matrix renormalization group approach (DMRG)
with the matrix-product state representation ($t$-MPS approach)~\cite{schollwock2005,schollwock2011,dolfi2014}.
It yields numerically-exact results on both equilibrium and out-of-equilibrium properties of low dimensional lattice models. The approach resorts on the Schmidt expansion of the many-body wave function and permits to reduce the Hilbert space to a finite, relevant subset, provided the entanglement entropy remains sufficiently small.
Owing to the area law~\cite{wolf2008,eisert2010}, it is optimal for 1D lattice models with a finite local Hilbert space in gapped phases, the entanglement of which remains finite in the thermodynamic limit.
It also applies to gapless phases, although with more stringent numerical parameters (high-filling cut-off and the bond dimension). 
To validate the accuracy of out results in all phases of the BH model, a systematic study of the effect of these parameters has been performed. 

\bigskip
\lettersection{Truncation of the local Hilbert space}
For the BH model considered in this work, the local Hilbert space is spanned by the Fock basis of number states, $\vert n_R\rangle$, where $n_R\in\mathbb{N}$, which is infinite. However, the probability distribution of the lattice-site occupation $n_R$ decays faster than exponentially in both the SF and MI phases. Accurate results can thus be obtained by cutting off the local Hilbert space to some value $n_{\textrm{\tiny max}}$.
It is important to note that, in some cases, the value of $n_{\textrm{\tiny max}}$ needs to be significantly much larger than the average filling $\qav{n}$ and its fluctuations.
This observation is consistent with analyses of truncated Bose-Hubbard models in quantum Monte Carlo simulations~\cite{kashurnikov1998zero}.

The SF meanfield regime, which corresponds to a high filling factor $\bar{n}$ and the gapless dispersion relation, has the most binding criteria. We found that a good estimator for $n_{\textrm{\tiny max}}$ is given by the condition
$1 - \sum_{n=0}^{n_{\textrm{\tiny max}}} P(n) \lesssim 10^{-2}$,
where $P(n)$ is the probability that $n$ bosons occupy a given lattice site.
In the SF meanfield regime, the probability distribution is nearly Poissonian,
$P(n) \simeq \bar{n}^n e^{-\bar{n}}/n!$. For instance, for the filling factor $\qav{n}=5$ used for the data of Fig.~\ref{fig:velocities_superfluid}, it yields $n_{\textrm{\tiny max}} \gtrsim 12$. 
For the strongly correlated SF regime at $\qav{n}=1$ considered for Fig.~\ref{fig:numerics}(a), the density fluctuations are significantly
suppressed and using the same condition as previously leads to $n_{\textrm{\tiny max}}=5$. For the MI phase at $\bar{n}=1$ and moderate values of
$U/J$ ($15 \geq U/J \geq u_\mathrm{c}$) considered for Fig.~\ref{fig:numerics}(b), we kept $n_{\textrm{\tiny max}}=5$. Deep in the MI phase 
($U/J \geq 15$), truncating the local Hilbert space to $n_{\textrm{\tiny max}}=2$, as used for Fig.~\ref{fig:numerics}(c) turns out to be sufficient.
Finally, the strongly interacting SF regime is the easiest case from a numerical point of view.
Owing to the low filling factor $\bar{n}<1$ and the large value of the interaction parameter $U/J$, the above condition also yields $n_{\textrm{\tiny max}}=2$, as used for Fig.~\ref{fig:velocities_g2_sir_sf}.
In all cases, we have checked that the numerics are converged for these values of $n_{\textrm{\tiny max}}$.

\bigskip
\lettersection{Bond dimension}
Within the MPS approach, the many-body state for a $M$-site lattice is represented in the tensor network form
\begin{equation}
\ket{\Psi} = \sum_{n_1, n_2, \ldots n_{M}} A^{n_1}[1] \, A^{n_2} [2] \ldots A^{n_{M}} [M]
 \ket{n_1, n_2, \ldots , n_{M}},
\end{equation}
where $n_j$ spans a local Hilbert space basis. For the BH model, it corresponds to a Fock basis truncated at $n_{\textrm{\tiny max}}$. 
For each value of $n_j$, the quantity $A^{n_j} [j]$ is a $\chi_{j-1}\times \chi_{j}$ matrix, where $\chi_{j}$ is the rank associated 
to the Schmidt matrix when applying the $j$-th singular value decomposition~\cite{schollwock2011}.
The bond dimension $\chi$ is defined as the maximum rank, $\chi = \mathrm{max}_j \left( \chi_j \right),~j \in [0 \ldots M]$. 
Note that for open-boundary conditions, the quantities $A^{n_1}[1]$ and $A^{n_{M}}[M]$ are actually
a row vector and a column vector, respectively, \ie\ $\chi_0 = \chi_{M} = 1$.

In the numerics, the maximum value of $\chi$ is chosen sufficiently large so that the truncation does not affect the results.
In practice, the calculations are run for several values of $\chi$ up to convergence of the correlation function $G_1(R,t)$ or $G_2(R,t)$. 
The required value of $\chi$ significantly depends on the regime and on the observable. In the following, we give the values used
for the final results presented in the paper.

For the SF meanfield regime [Figs.~\ref{fig:velocities_superfluid}(a) and \ref{fig:velocities_g1_superfluid}], we used the values 
$\chi = 300$ and $\chi=450$ for the $G_2$ and $G_1$ functions, respectively. The bond dimension used for $G_1$ is higher than the one 
for $G_2$ due to the long-range phase correlations already present at equilibrium.
For the SF strongly correlated regime at $\qav{n}=1$ [Fig.~\ref{fig:numerics}(a)], we used $\chi = 300$ for both correlation functions. A similar
value of $\chi$ was considered for moderate values of $U/J$ in the MI phase at $\bar{n}=1$ [Fig.~\ref{fig:numerics}(b)]. Deep in the MI phase 
[Fig.~\ref{fig:numerics}(b)], the bond dimension can be significantly decreased and we consider $\chi = 100$. Finally, in the SF 
strongly interacting regime at $U/J=50$, we found that the value $\chi = 100$ is enough.

\section{One-body correlation function $\mathbf{G_1(R,t)}$ in the meanfield regime}
\label{g1_mf}
\begin{figure}[h!]
\includegraphics[scale=0.31]{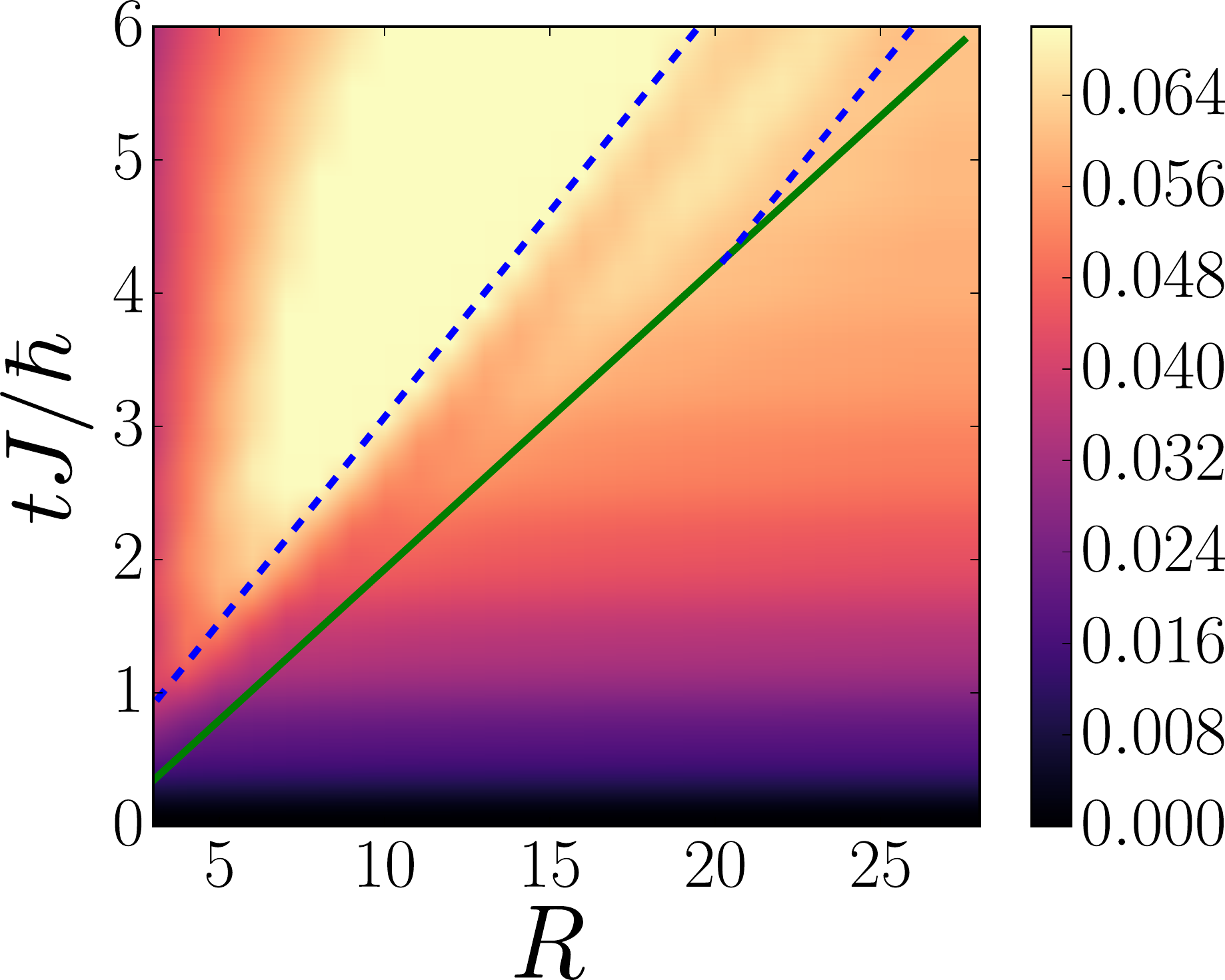}
\caption{\label{fig:velocities_g1_superfluid}
Spreading of the one-body correlation function $G_1(R,t)$ for a global quench in the SF meanfield regime from $(U/J)_0 = 0.2$ to $U/J = 0.1$ and $\bar{n}=5$. The solid-green and dashed-blue lines are fits to the CE and maxima, respectively. 
}
\end{figure}

In the analysis of the SF meanfield regime reported in the main paper, we focused on the two-body correlation function $G_2(R,t)$.
We have also studied the one-body correlation $G_1(R,t)$ using the same $t$-MPS simulations.
We found that the dynamics of the $G_1$ function shows a spike-like structure, similar to that found for the $G_2$ function. The values of the correlation edge ($\VCE$) and maxima ($\Vm$) velocities agree with those found for the $G_2$ function within less than $10\%$.
Figure~\ref{fig:velocities_g1_superfluid} shows an example, for the quench from $(U\qav{n}/J)_0 = 1$ to $U\qav{n}/J = 0.5$, and $\bar{n}=5$. The fits to the correlation edge and to the maxima yield the velocities
$\VCE = (4.4 \pm 0.3)\, J/\hbar$ and $\Vm = (3.3 \pm 0.2)\, J/\hbar$,
in excellent agreement with the corresponding values found from the dynamics of the $G_2$ function, see Fig.~\ref{fig:velocities_superfluid}(b).

The agreement between the spreading velocities for different correlation functions was found in all regimes, see for instance Figs.~\ref{fig:numerics}(d1) and (d2).
It is consistent with the prediction that these velocities are characteristic of the excitation 
spectrum and not on the details of the correlation function~\cite{cevolani2018}.
Note, however, that the full space-time dependence of the signal depends on the correlation function.
In general, we found that the signal for $G_1$ is less sharp than for $G_2$. This may be attributed to the long-range phase correlations present in the initial state, which blur the correlation function~\cite{bravyi2006}.

\section{Mapping on the 1D Lieb-Liniger model}
\label{mapping}

In the long-wave length regime, the lattice discretization of the Bose-Hubbard (BH) may be disregarded.
The BH model then maps onto the continuous-space Lieb-Liniger (LL) model,
\begin{equation}
\label{LLmodel}
\hat{H} = \frac{\hbar^2}{2m} \left[ - \sum_{i=1}^{N} \frac{\partial^2}{\partial x_i^2} + c\sum_{i\neq j} \delta(x_i - x_j) \right].
\end{equation}
It describes a one-dimensional gas of $N$ bosons of mass $m$ with contact interactions, 
characterized by the interaction strength $c>0$.
The correspondance between the parameters of the BH and LL models is found by discretizing the LL model, Eq.~(\ref{LLmodel}), on the length scale defined by the lattice spacing $a$. It yields
$J = \hbar^2/2ma^2$ and $U = \hbar^2 c /ma$.
The density of the LL model is $\rho \equiv N/L = \qav{n}/a$, where $\qav{n}$ is the number of bosons per lattice site (filling) and $L$ is the system size.

The LL Hamiltonian is exactly solvable by Bethe ansatz~\cite{lieb1963a,lieb1963b}.
All the thermodynamic quantities at zero temperature can be written as universal functions of
the Lieb-Liniger parameter $\gamma=c/\rho$
and the dimensionless quantity
$e(\gamma)=E_0/Nn^2$, where $E_0$ is the ground state energy.
For instance, the macroscopic sound velocity~\cite{lieb1963b} reads as
\begin{equation}
v_\textrm{s} \equiv \sqrt{\frac{L}{m\rho}\left.\dfrac{\partial^{2}E_0}{\partial L^{2}}\right|_{N,S}}
=\dfrac{\hbar\rho}{m}\sqrt{3e(\gamma)-2\gamma e'(\gamma)+\dfrac{1}{2}\gamma^{2}e''(\gamma)}.
\end{equation}
Using the small $\gamma$ expansion, $e(\gamma)=\gamma\left[1-(4/3\pi)\sqrt{\gamma}\right]$, one then finds
\begin{equation}
v_\textrm{s}=\dfrac{\hbar \rho}{m}\sqrt{\gamma}\big(1-\sqrt{\gamma}/4\pi\big),
\end{equation}
valid in the weakly-interacting regime, $\gamma \ll 1$.
Finally, using the correspondance between the parameters of the BH and LL models,
one finds
\begin{equation}
\Vs \equiv v_\textrm{s}/a= \frac{2J\qav{n}}{\hbar}\sqrt{\gamma}\big(1-\sqrt{\gamma}/4\pi\big)
\end{equation}
and $\gamma=U/2J\qav{n}$.

\section{Two-body correlation function $\mathbf{G_2(R,t)}$ in the Mott-insulating phase}
\label{spa}
In order to explain the suppression of the twofold structure for the two-body correlations deep in the Mott insulator phase (MI; $U\gg J$ and $\qav{n}=1$), we compute the function $G_2(R,t)$, working 
along the lines of Ref.~\cite{barmettler2012}. Considering the manifold of doublon-holon pairs and mapping the resulting 
Hamiltonian into a fermionic one, the two-body correlation function may be written as

\begin{equation}\label{eq:MIG2}
G_2(R,t) \simeq -2 \big( |g_2(R,t)|^2 +|\bar{g}_2(R,t)|^2 \big),
\end{equation}

with

\begin{eqnarray}
&& g_2(R,t) \sim \frac{J}{U}\frac{R}{t} \int_{-\pi}^{+\pi} \frac{\textrm{d}k}{2 \pi} \Big\{\e^{i \left(2E_kt + kR\right)} + \e^{i \left(2E_kt - kR \right)} \Big\},
\label{g2} \\
\nonumber \\
&& \bar{g}_2(R,t) \sim \left(\frac{J}{U} \right)^2 \int_{-\pi}^{+\pi} \frac{\textrm{d}k}{2 \pi} \sin^2(k) \Big\{\e^{i \left(2E_kt - kR\right)} + \e^{-i \left(2E_kt + kR \right)} \Big\}
\label{g2_bar}
\end{eqnarray}

and the excitation spectrum is $2E_k \simeq \sqrt{ \left[ U - 2J(2\bar{n} \! + \! 1)\cos(k)\right]^2 + 16 J^2 \bar{n}(\bar{n} \! + \! 1)\sin^2(k)}$,
see Eq.~\eqref{mott_disp_rel_fermionization}.

\bigskip
\lettersection{Quench deep into the Mott insulator phase}
For a quench, very deep in the MI phase, $U \gg J$, the second right-hand-side term in Eq.~(\ref{eq:MIG2}) is much smaller
than the first one and the former can be neglected. Using Eq.~(\ref{g2}), it yields explicitly
for $G_2(R,t) \simeq -2|g_2(R,t)|^2$,

\begin{equation}\label{eq:MIG2bis}
G_2(R,t) \sim - 2 \left(\frac{J}{U}\right)^2 \left(\frac{R}{t}\right)^2 \left| \int_{-\pi}^{\pi} \frac{\textrm{d}k}{2 \pi}
\Big\{ \e^{i \left(2E_kt + kR\right)} + \e^{i \left(2E_kt - kR \right)} \Big\} \right|^2
\end{equation}

Moreover, the excitation spectrum may be expanded in powers of $J/U$. Up to first-order, it yields $2E_k \simeq U-2J(2\bar{n}+1)\cos(k)$.
The gap term $\e^{iUt}$ can then be factorized in the two terms under the integral in Eq.~(\ref{eq:MIG2bis}) and disappears due to the square modulus.
Introducing the effective excitation spectrum $ 2\tilde{E}_k = -2J(2\bar{n}+1)\cos(k)$, we then find
$G_2 \simeq -2 \vert g_2(R,t)\vert^2$ with  

\begin{equation}\label{eq:MIG2bis2}
g_2(R,t) \sim \frac{J}{U} \frac{R}{t} \int_{-\pi}^{\pi} \frac{\textrm{d}k}{2 \pi}
\Big\{ \e^{i \left(2\tilde{E}_kt + kR\right)} + \e^{i \left(2\tilde{E}_kt - kR \right)} \Big\}.
\end{equation}

The integral may be evaluated using the stationary phase approximation.
In the infinite time and distance limit along the line $R/t=\mathrm{cst}$, the integral in Eq.~\eqref{eq:MIG2bis2} is dominated by
the momentum contributions with a stationary phase (sp), \ie\ $\partial_k (2\tilde{E}_k t \pm kR) = 0$ or, equivalently,
$2\tildeVg (\ksp) = \pm R/t$ where $\tildeVg=\partial_k \tilde{E}_k$ is the group velocity of the effective excitation spectrum.
Since the latter is upper bounded by the value $\tildeVg^* = \mathrm{max}(\tildeVg) = J(2\bar{n}+1)$, it has a solution only for $R/t < 2\tildeVg^*$.
We then find

\begin{equation}
\label{g2_ksp}
g_2(R,t) \sim \frac{J}{U} \frac{\tildeVg(\ksp)}{\left(|\partial_k^{2}\tilde{E}_{\ksp}| t\right)^{1/2}} \left[
\cos\left(2\tilde{E}_{\ksp}t - \ksp R + \sigma \frac{\pi}{4} \right) + i\sin \left(2\tilde{E}_{\ksp}t - \ksp R + \sigma\frac{\pi}{4} \right) \right].
\end{equation}

with $\sigma = \mathrm{sgn} \left(\partial_k^2 \tilde{E}_{\ksp} \right)$. For both the real and imaginary parts of $g_2(R,t)$, the
correlations are activated ballistically at the time $t=R/2\tildeVg^*$. It defines a linear correlation edge (CE) with velocity $\VCE = 
2\tildeVg^*$. In addition, Eq.~\eqref{g2_ksp} also yields a series of local maxima, defined by the equation $2\tilde{E}_{\ksp}t - \ksp R = \mathrm{cst}$. In the vicinity of the CE cone,
these maxima (m) propagate at the velocity $\Vm = 2\tildeVphi^* = 
2\tilde{E}_{k^*}/k^*$, \ie\ twice the phase velocity at the maximum of the group velocity, $k^*$.

\begin{figure*}[h!]
\includegraphics[width = \columnwidth]{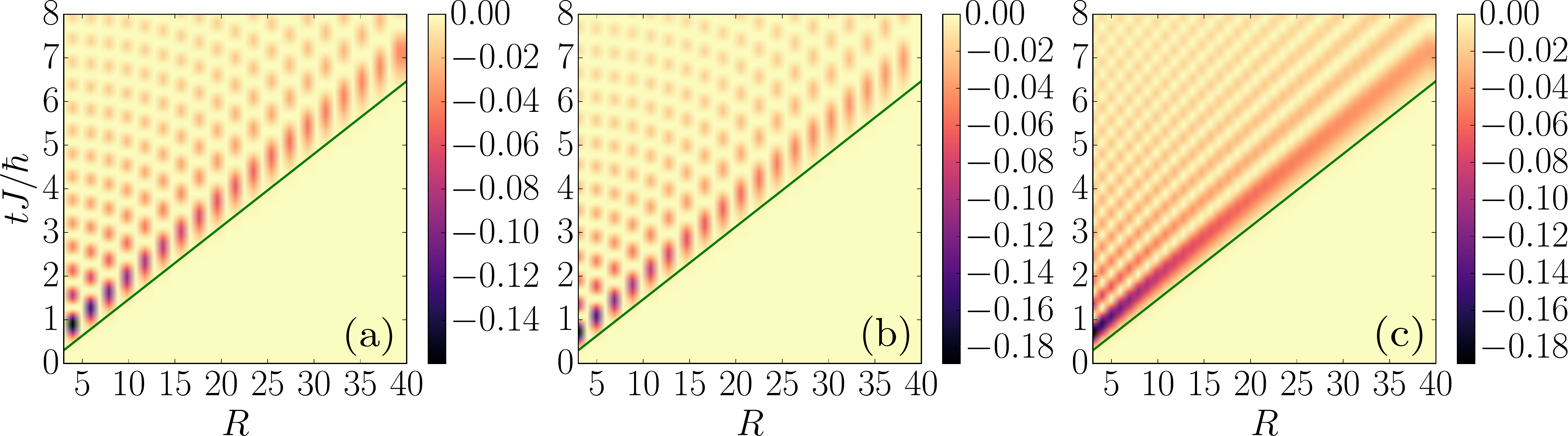}
\caption{\label{fig:g2_analysis}
Analysis of the space-time correlation pattern of $G_2(R,t)$ via $g_2(R,t)$ [see Eq.~\eqref{eq:MIG2bis2}] at $\bar{n}=1$
for a global quench confined deep into the Mott-insulating phase starting from a pure Mott state $(U/J)_0 \rightarrow \infty$.
Analytical expression, owing to prefactors, of (a)~ $-\Re^2 \left[g_2(R,t) \right]$ (b)~ $-\Im^2 \left[g_2(R,t) \right]$ (c)~ sum
of the two contributions shown at Fig.~(a) and (b). The solid green line corresponds to the theoretical CE velocity characterized by
$2\tildeVg^* = 2J(2\bar{n}+1)$. On Fig.~(c), the first extremum propagates with the same velocity as the one associated to the CE. 
}
\end{figure*}

Hence, the real and imaginary parts of $g_2(R,t)$ both display a twofold structure with a CE velocity $2\tildeVg^* = 2J(2\bar{n}+1)$ and  a velocity of the maxima $2\tildeVphi^* = 0$, as shown on Figs.~\ref{fig:g2_analysis}(a) and (b).
In contrast, $G_2(R,t)$, does not display the twofold structure. This is because it is the sum of the squares of the two latter contributions [see Eq.~\eqref{g2_ksp}], which are shifted by half a period and cancel each other.
It thus gives a single cone structure, characterized by the sole CE velocity $2\tildeVg^*$, as shown on Fig.~\ref{fig:g2_analysis}(c).

\lettersection{Quench into the Mott insulator phase for moderate $U/J$}
For moderate values of $U/J$, still in the MI phase, the second term in the right-hand-side of Eq.~(\ref{eq:MIG2}), 
$|\bar{g}_2(R,t)|^2$, becomes relevant. Using again the stationary-phase approximation for $\bar{g}_2(R,t)$, we find
\begin{equation}
\bar{g}_2(R,t) \sim \left(\frac{J}{U}\right)^2 \frac{\sin^2(\ksp)}{\left(|\partial_k^{2}E_{\ksp}| t\right)^{1/2}} 
\cos\left(2E_{\ksp}t - \ksp R + \sigma' \frac{\pi}{4} \right) 
\end{equation}
with $\sigma' = \mathrm{sgn}\left(\partial_k^2 E_{\ksp} \right)$ and $E_k$ the excitation spectrum given 
at Eq.~\eqref{mott_disp_rel_fermionization}.
Using the same argument as above, we find that $\bar{g}_2(R,t)$ 
shows a twofold structure characterized by, now, the CE velocity $2\Vg^* = 2 \mathrm{max}\left(\partial_k E_k \right)$ but the velocity of the maxima
$2V_\varphi^* = 2E_{k^*}/k^* \neq 0$.
Since there is a single contribution here, the quantity $\vert\bar{g}_2(R,t)\vert^2$ displays a twofold structure with the same characteristic velocities. More precisely, both the length and time scales of the oscillations are divided by two but the velocities are not affected.

For a quench into the MI phase at a moderate value of $U/J$, both $|g_2(R,t)|^2$ and $|\bar{g}_2(R,t)|^2$ contribute to the two-body correlation function $G_2(R,t)$.
While the $|g_2(R,t)|^2$ contribution is characterized by the sole CE velocity $2\Vg^*$,
the $|\bar{g}_2(R,t)|^2$ contribution provides the double structure observed on $G_2$ for $6<U/J<10$ in the $t$-MPS calculations. 

\end{document}